\documentclass[a4wide,12pt]{article}
\usepackage{graphics}
\usepackage{amsmath,amssymb}

\def\ve{\varepsilon}
\def\half{\frac{1}{2}}
\def\quarter{\frac{1}{4}}
\def\be{\begin{equation}}
\def\ee{\end{equation}}
\def\bea{\begin{eqnarray}}
\def\eea{\end{eqnarray}}
\def\beax{\begin{eqnarray*}}
\def\eeax{\end{eqnarray*}}
\begin{document}
\setcounter{footnote}{1}
\renewcommand{\thefootnote}{{\fnsymbol{footnote}}}
\title{Interface between Hermitian and non-Hermitian Hamiltonians in a model calculation}
\author{H. F. Jones\footnote{e-mail: h.f.jones@imperial.ac.uk}\\
Physics Department, Imperial College, London SW7 2AZ, UK}
\maketitle
\begin{abstract}
We consider the interaction between the Hermitian world, represented by a real delta-function
potential $-\alpha\delta(x)$, and the non-Hermitian world, represented by a PT-symmetric pair
of delta functions with imaginary coefficients $i\beta(\delta(x-L)-\delta(x+L))$. In the context
of standard quantum mechanics, the effect of the introduction of the imaginary delta functions on
the bound-state energy of the real delta function and its associated
wave-function is small for $L$ large. However, scattering from the combined potentials does not
conserve probability as conventionally defined. Both these problems can be studied instead in the
context of quasi-Hermiticity, whereby quantum mechanics is endowed with a new metric $\eta$, and consequently
a new wave-function $\Psi(x)$, defined in terms of the original wave-function $\psi(x)$ by means of $\eta$.
In this picture, working perturbatively in $\beta$,  the bound-state wave-function is actually unchanged
from its unperturbed form for $|x|\ll L$. However, the scattering wave-function, for $|x|\gg L$, is changed
in a significant manner. In particular there are incoming and outgoing waves on both sides of the potential.
One can then no longer talk in terms of reflection and transmission coefficients, but the total right-moving
flux is now conserved.
\end{abstract}
\section{Introduction}
Since the resurgence of interest in non-Hermitian Hamiltonians initiated by the paper of
Bender and Boettcher\cite{BB} the subject has gone through several stages. First there
was the search for soluble non-Hermitian Hamiltonians with real spectra (see, e.g. \cite{GZ}).
The next stage concerned the introduction of a new Hilbert-space metric in order to obtain
positive probabilities and so regain a physical interpretation of the theory. This was first
done in $PT$-symmetric theories\cite{BBJ} by introducing the so-called $C$ operator to form the metric
$CPT$. A more general framework was formulated by Mostafazadeh\cite{AM-metric}, which among other things
established the connection to earlier work by Scholtz et al.\cite{Hendrik} and showed\cite{AM-h} that such a Hamiltonian $H$ was related by a similarity transformation to an equivalent Hermitian Hamiltonian $h$. Subsequent work showed how the metric $\eta$ could be constructed, sometimes exactly, but more typically in perturbation theory for a variety of models\cite{HFJ-ix3, AM-ix3, -x4}.

More recently some attention
has been given to situations where a non-Hermitian system interacts with the world of Hermitian quantum mechanics.
For example, Ref.~\cite{QB} examined a non-Hermitian analogue of the Stern-Gerlach experiment in which the role of the intermediate inhomogeneous magnetic field flipping the spin is taken over by an apparatus described by a non-Hermitian Hamiltonian. This type of set-up has been further elaborated by Assis and Fring\cite{Andreas} and Guenther al.\cite{Uwe}, and Mostafazadeh\cite{AM-QB} has emphasized that the effect relies on non-unitarity in
some guise or another (see also \cite{comment}). The subject has also taken an experimental turn in the form
of optical lattices whose refractive index can be tailored to make them $PT$-symmetric (see, for example, \cite{Christos}).

An earlier attempt at understanding the conceptual issues involved in scattering from a non-Hermitian
potential was given in Ref.~\cite{HFJ-1df}. There we primarily considered a single delta-function at the origin
with a complex coefficient. Here unitarity is not conserved within the framework of conventional quantum mechanics, but if one instead constructs the metric $\eta$ and the corresponding transformed wave-function $\Psi$ (see Eq.~(\ref{Psi}) below) unitarity is restored but at the price of a drastic change in the physical picture,
whereby there are now incoming and outgoing waves on both sides of the potential. A somewhat less drastic, but
still significant, change is found by Znojil\cite{MZ-scatt} in a discretized model of scattering, where the
metric does not mix up incoming and outgoing waves, but instead changes the normalization of the flux on either side of the scattering centre. In Ref.~\cite{HFJ-1df} we touched on the potential model that we address in the
present paper, but only in the context of conventional quantum mechanics. On the basis of those calculations
we speculated that the treatment of bound-state problems should be essentially unaffected by the introduction of distant non-Hermitian scattering potentials, and that it is only when the system
physically interacts with those potentials that a drastic change in the formalism
is required, if indeed the non-Hermitian potentials are regarded as
fundamental rather than effective.

In the present paper we return to that model, which we are now able to treat in the quasi-Hermitian picture
as well, thanks to a general prescription due to Mostafazadeh\cite{AM-eta} for calculating $\eta$ perturbatively in the case of a superposition of delta functions. The model is introduced in Section 2, where it is treated
in the framework of conventional quantum mechanics. As was already seen in Ref.~\cite{HFJ-1df}, the bound-state
energy and wave-function are affected only by exponentially small terms when $L\to\infty$, but the scattering
does not preserve unitarity, with $R+T\ne 1$. In Section 3 we instead treat the problem in the
quasi-Hermitian framework, calculating the metric to first order in $\beta$ but to all orders in $\alpha$.
For the bound state we find that the new wave-function $\Psi$ is actually equal to the old wave-function
$\psi$ in the absence of the non-Hermitian part of the Hamiltonian. For the scattering problem we find the
same type of wave-function that we previously found for the single complex delta function, whereby $\Psi$ contains an incoming wave on the right-hand side, in contrast to $\psi$. Finally, in Section 4, we discuss the significance of these results.

\section{Delta-Function Model}

The Hamiltonian we analyze in this paper has a potential consisting of three delta functions:
\bea\label{potential}
H=p^2-\alpha\delta(x) +i\beta(\delta(x-L)-\delta(x+L)).
\eea
The first component of the potential, $-\alpha\delta(x)$, is a delta function based on the origin with a real,
negative coefficient. When $\beta=0$ this Hermitian part of the potential supports a single bound
state with energy $E=-\kappa^2$, where $\kappa=\half\alpha$. The second component proportional to
$\beta$ consists of two delta functions based at $x=\pm L$ with imaginary coefficients $\pm i\beta$,
designed to be $PT$-symmetric and have real energy eigenvalues. What is at issue is how the introduction
of this non-Hermitian piece of the Hamiltonian affects both the scattering wave-function and the
bound-state energy and wave-function. Of particular interest is the case when $L$ is large, i.e.
when the non-Hermitian pieces are distant from the Hermitian potential based at the origin.
\subsection{Bound State}
Let us start by setting $\beta=0$. As already stated, there is then a single bound state, with (unnormalized) wave-function
\bea\label{unp}
\psi=e^{-\kappa |x|},
\eea
where $\kappa=\half\alpha$ in order to satisfy the continuity condition
\bea\label{cont1}
\psi'(0_+)-\psi'(0_-)=\alpha \psi(0).
\eea

Let us now repeat this calculation for $\beta\ne0$. The bound-state wave-
function then has the ($PT$-symmetric) form
\bea
\psi(x)=\left\{\begin{array}{lc}
W e^{\kappa x}& \;\; x<-L \\
U e^{\kappa x}+V e^{-\kappa x} &\;\; -L<x<0 \\
U^* e^{-\kappa x}+V^* e^{\kappa x} &\;\; 0<x<L \\
W^* e^{-\kappa x}& \;\; L<x
\end{array}\right.
\eea
with $\psi(-x)=\psi^*(x)$.
Applying the continuity conditions  $\psi(-L_+)=\psi(-L_-)$,
$\psi'(-L_+)-\psi'(-L_-)=-i\beta \psi(-L)$ at $x=-L$ we find
\bea
U&=&(1-i\tilde{\beta}) W\nonumber\\
V&=& i\tilde{\beta}e^{-2\kappa L} W,
\eea
where $\tilde{\beta}=\beta/(2\kappa)$.
The continuity conditions at $x=0$ are
\bea
U+V&=&U^*+V^*\nonumber\\
\kappa[(U+U^*)-(V+V^*)]&=&\alpha(U+V).
\eea
The first of these gives
\bea
\left[(1-i\tilde{\beta})+i\tilde{\beta}e^{-2\kappa L}\right] W=
\left[(1+i\tilde{\beta})-i\tilde{\beta}e^{-2\kappa L}\right] W^*,
\eea
which can be satisfied by taking
\bea
W=(1+i\tilde{\beta})-i\tilde{\beta}e^{-2\kappa L},
\eea
up to an overall real normalization constant.
On substitution into the second continuity condition we find, after some algebra,
the eigenvalue equation
\bea
2\kappa\left[1+\tilde{\beta}^2(1-e^{-4\kappa L})\right]=\alpha\left[1+\tilde{\beta}^2(1-e^{-2\kappa L})^2\right],
\eea
which, after some additional manipulation, can be written as
\bea\label{alpha_tilde}
\tilde{\alpha}=1+\frac{(1+\tilde{\alpha})\tilde{\beta}^2 e^{-2\kappa L} (1-e^{-2\kappa L})}
{1+\tilde{\beta}^2(1-e^{-2\kappa L})},
\eea
where $\tilde{\alpha}=\alpha/(2\kappa)$. In this form it is clear that (i) $\kappa\to \half\alpha$ as $L\to\infty$, and (ii) the first correction
to $\kappa$ is of order $\beta^2$. It is interesting that the wave-function does not depend
explicitly on $\alpha$, only through the relation between $\kappa$ and $\alpha$.

\subsection{Scattering Wave-Functions}
Again let us start with $\beta=0$. The scattering wave-function, in the situation where a plane wave comes in from the left and is either
reflected or transmitted at the delta-function potential, is of the form
\bea
\psi(x)=\left\{\begin{array}{ll} A e^{ikx}+ B e^{-ikx}& \;\; x<0 \\ e^{ikx}&\;\; x>0 \end{array}\right.
\eea
The coefficients $A$ and $B$ are determined by the continuity condition (\ref{cont1}) as
\bea
A&=&1-i\hat\alpha\nonumber\\
B&=&i\hat\alpha,
\eea
where $\hat{\alpha}=\alpha/(2k)$,
giving reflection ($R$) and transmission ($T$) coefficients
\bea\begin{array}{lrl}
T=&1/|A^2|&=1/\left(1+\hat{\alpha}^2\right)\\ &&\\
R=&|B^2|/|A^2|&= \hat{\alpha}^2/\left(1+\hat{\alpha}^2\right).\end{array}
\eea
Since the potential for $\beta=0$ is real, the scattering is unitary, with $R+T=1$.

For $\beta\ne 0$ the wave-function has the general form
\bea
\psi(x)=\left\{\begin{array}{lc}
A e^{ikx}+B e^{-ikx}& \;\; x<-L \\
C e^{ikx}+D e^{-ikx} &\;\; -L<x<0 \\
E e^{ikx}+F e^{-ikx} &\;\; 0<x<L \\
e^{ikx}&\;\; L<x
\end{array}\right.
\eea
Working from the right, we first apply the continuity conditions at $x=L$, to obtain
\bea
E&=&1-\hat\beta\nonumber\\
F&=&\hat\beta e^{2ikL},
\eea
where $\hat\beta=\beta/(2k)$. Then, applying the continuity conditions at $x=0$, we obtain
\bea
C&=&(1-i\hat\alpha)(1-\hat\beta) +i\hat\alpha\hat\beta e^{2ikL}\nonumber\\
D&=&(1+i\hat\alpha)\hat\beta e^{2ikL} +i\hat\alpha(1-\hat\beta).
\eea
Finally, applying the continuity conditions at $x=-L$ we obtain, after some algebra,
\bea
A&=&(1-{\hat\beta}^2)(1-i\hat\alpha)-2i\hat\alpha{\hat\beta}^2  e^{2ikL}
+{\hat\beta}^2(1+i\hat\alpha) e^{4ikL}\\
B&=& \hat\beta(1-\hat\beta)[(1+i\hat\alpha)e^{2ikL}-(1-i\hat\alpha)e^{-2ikL}]+i\hat\alpha[{\hat\beta}^2+(1-\hat\beta)^2].\nonumber
\nonumber
\eea
Note that $A$ only involves ${\hat\beta}^2$, whereas $B$ involves $\hat\beta$ linearly. The coefficients
are at most linear in $\alpha$. As expansions in $\beta$ we have
\bea
A&=&1-i\hat\alpha+O(\beta^2)\\
B&=&i\hat\alpha+2i\hat\beta[\sin{2kL}-2\hat{\alpha}\sin^2{kL}]+O(\beta^2).\nonumber
\eea
Thus the transmission and reflection coefficients are
\bea
\left.\begin{array}{ll}T=&1/(1+\hat\alpha^2) \\ R=&(\hat\alpha^2+4\hat\alpha\hat\beta\sin{2kL})/(1+\hat\alpha^2)\end{array} \right\}+O(\beta^2, \alpha^2\beta).
\eea
Clearly unitarity, as conventionally defined, is violated in this process.
\section{Quasi-Hermitian Analysis}
The Hamiltonian has been specifically constructed to be $PT$ symmetric. In such cases, as discussed
in the introduction, we can in principle introduce a positive-definite metric operator\cite{AM-metric, usQ} $\eta=e^{-Q}$
with respect to which $H$ is quasi-Hermitian: $H^\dag=\eta H \eta^{-1}$. Observables are those represented
by quasi-Hermitian operators $A$ such that $A^\dag=\eta A \eta^{-1}$. The original position operator $x$ does
not fall into this category: instead the observable of position is $X\equiv \rho x\rho^{-1}$, where
$\rho=\eta^\half=e^{-\half Q}$. Consequently\cite{AM-Psi,MB}, in this picture the relevant wave-function
is not $\psi(x)\equiv\langle x|\psi\rangle$, but\vspace{-2mm}
\bea\label{Psi}
\Psi(x)\equiv \langle x|\Psi\rangle=\langle x|\rho|\psi\rangle={\tiny\int} \rho(x,y)\psi(y)dy.
\eea
Let us therefore attempt to construct $\Psi$ for the present Hamiltonian. Unfortunately this cannot be
done exactly, but Mostafazadeh\cite{AM-eta} has devised a general method for constructing a series expansion for $\eta$ in the coupling constants of a series of delta functions. This method is based on expressing the
condition of quasi-Hermiticity of the Hamiltonian as a partial differential equation for $\eta(x,y)$, and converting it into an integral equation that can be solved iteratively.

In some more detail, for the Hamiltonian $H= p^2+V(x)$, the integral equation for $\eta(x,y)$
takes the form
\bea\label{LS}
\eta(x,y)=u(x,y)+(K\eta)(x,y),
\eea
where $u(x,y)\equiv u_+(x-y)+u_-(x+y)$ is the general solution of the differential equation
$(-\partial_x^2+\partial_y^2)u(x,y)=0$, and $K$ is defined by
\bea
(K\eta)(x,y)=\left(\int^y dr\ V(r)\int^{x+y-r}_{x-y+r} ds +\int^x ds\ V^*(s)\int^{x+y-s}_{y-x+s} dr\right) \eta(s,r)
\eea
In Ref.~\cite{AM-eta} the equation (\ref{LS}) was written in the form
\bea\label{LS2}
\eta=(1-K)^{-1}u,
\eea
which in principle can be expanded in $K$, i.e. as a simultaneous expansion in the coefficients
of the delta functions. However, in the present case we need to do something slightly different,
because we are thinking of $\beta$ as a perturbative parameter, but not $\alpha$. Thus we need to split $K$ up
into $K=K_\alpha+K_\beta$. Then we write Eq.~(\ref{LS}) as
\bea
\eta=u+(K_\alpha+K_\beta)\eta.
\eea
Now for $\beta=0$ the Hamiltonian is Hermitian, so we want $\eta(x,y)=\delta(x-y)$. This means
that to order $\beta$ we should take $u=(1-K_\alpha)\delta$. In principle we could also add to $u$ a term $\beta w$, where $w$ is a solution of the
homogenous equation $(\partial_x^2-\partial_y^2)w(x,y)=0$. However, it turns out that such a term is not required. So
\bea
\eta=(1-K_\alpha)\delta+(K_\alpha+K_\beta)\eta,
\eea
so that
\bea
(1-K_\alpha)\eta=(1-K_\alpha)\delta+K_\beta \eta ,
\eea
with solution
\bea
\eta=\delta+\frac{1}{1-K_\alpha} K_\beta\eta\ .
\eea
To $O(\beta)$, which is as far as we will take the calculation, this reduces to
\bea\label{K}
\eta&=&\delta+\frac{1}{1-K_\alpha} K_\beta\delta\nonumber\\ \cr
&=& \delta+(1+K_\alpha+K_\alpha^2+\dots)K_\beta\delta\ .
\eea
Since we would like to treat $\beta$ perturbatively, but not $\alpha$, it is extremely fortunate that the higher powers
of $K_\alpha$ in this equation do not in fact contribute, as is shown in Section 3.2.

\subsection{Effect of $K_\beta$}
As Mostafazadeh has shown\cite{AM-eta} (with $m=\half$, $\hbar=1$), the action of $K_\beta$ on the delta function $\delta(x-y)$ is
\bea
(K_\beta \delta)(x,y)=\half i\beta[\theta(x+y-2L)-\theta(x+y+2L)]\ve(y-x),
\eea
where $\ve(z)$ is the sign function $\ve(z)\equiv {\rm sgn}(z)$.
Thus, to order $\beta$  we have
\bea\label{eta}
\eta(x,y)
&=&\delta(x-y)+\left\{\begin{array}{lc}
0& \;\; x+y<-2L \\
\half i\beta\ve(x-y) & \;\; -2L<x+y<2L \\
0& \;\; 2L<x+y \end{array}
\right.\nonumber\\
&\equiv&\delta(x-y)-\beta Q_1(x,y)
\eea
\subsubsection{Bound-state wave function}
Let us now calculate the effect of $\rho(x,y)= \delta(x-y)-\half\beta Q_1(x,y)$ on the bound-state wave-function
$\Psi(x)$.
From Eq.~(\ref{Psi}) it is given, to this order, by
\bea
\Psi(x)&=&\psi(x)+\frac{1}{4} i\beta\int_{-2L-x}^{2L-x}\ve(x-y)\psi(y)dy\nonumber\\
&=&\psi(x)-\frac{1}{4} i\beta\left[\int_x^{2L-x}\psi(y) dy -\int^x_{-2L-x} \psi(y) dy\right],
\eea
for $0< x < L$.
Here, since we are working to $O(\beta)$, we can take $\psi(y)=e^{-\kappa|y|}$ in the integrands. Then
\bea\label{Psibound}
\Psi(x)&=&(1+i\tilde\beta e^{-2\kappa L})e^{-\kappa x}-i\tilde\beta  e^{-2\kappa L}e^{\kappa x}\nonumber\\
&&-\frac{1}{4} i\beta \left[\int_0^{2L-x} e^{-\kappa y}dy -2\int_0^x e^{-\kappa y} dy
- \int_{-2L-x}^{0}  e^{\kappa y} dy\right]\\
&=&\left(1+\half i\tilde\beta e^{-2\kappa L}\right)e^{-\kappa x}-\half i\tilde\beta  e^{-2\kappa L}e^{\kappa x}+i\tilde\beta(1-e^{-\kappa x})+O(\beta^2).\nonumber
\eea
The last term is not something we expect at all, but before jumping to conclusions we should await the calculation
of the $K_\alpha K_\beta$ contribution, which, because it involves $\alpha$ in the combination
$\tilde\alpha\equiv\alpha/(2\kappa)=1+O(\beta^2)$, is of exactly the same order.

\subsubsection{Scattering wave-function}
We are now concerned with the scattering wave-function for large $|x|$.
For $x>0$ it is actually sufficient to take $x>3L$ (so that $2L-x<-L$), at which point the wave-function, $\Psi_>$
settles down to its asymptotic form, namely
\bea
\Psi_>(x)&=&e^{ikx}+\frac{1}{4} i\beta\int_{-2L-x}^{2L-x}\ve(x-y)[(1-i\hat\alpha)e^{iky}+i\hat\alpha e^{-iky}]dy
\nonumber\\
&=&e^{ikx}+\frac{1}{4} i\beta\int_{-2L-x}^{2L-x}[(1-i\hat\alpha)e^{iky}+i\hat\alpha e^{-iky}] dy\\
&=&(1-\hat\alpha\hat\beta\sin{2kL})e^{ikx}+i\hat\beta(1-i\hat\alpha)\sin{2kL}\ e^{-ikx}+O(\beta^2)\ .\nonumber
\eea
For $x<-3L$ (so that $-2L-x>L$) we instead get
\bea
\Psi_<(x)&=&Ae^{ikx}+Be^{-ikx}+\frac{1}{4} i\beta\int_{-2L-x}^{2L-x}\ve(x-y)e^{iky}dy\nonumber\\
&=&Ae^{ikx}+Be^{-ikx}-\frac{1}{4} i\beta\int_{-2L-x}^{2L-x}e^{iky} dy\\
&=&(1-i\hat\alpha)e^{ikx}+i[\hat\alpha+\hat\beta\sin{2kL}- 4\hat\alpha\hat\beta\sin^2{kL})]e^{-ikx}+O(\beta^2)\ .\nonumber
\eea
Note that the physical picture has completely changed, because we have incoming and outgoing waves on
both sides. This is the rather drastic modification noted in Ref.~\cite{HFJ-1df}. Some such modification
at infinity is clearly necessary if we are to restore unitarity in this picture. In a recent discretized
model of scattering studied by Znojil\cite{MZ-scatt} the modification is instead a change in the normalization of
the fluxes on either side of the scattering centre. In the present model we can check unitarity by comparing the net
right-moving fluxes $\Phi$ on each side. Unitarity is indeed restored to this order, because
\bea\label{firstfluxes}
\Phi_>=\Phi_<=1-2\hat\alpha\hat\beta\sin{2kL}+O(\alpha^2\beta,\beta^2).
\eea
Note that we are only allowed to calculate the fluxes using the standard formula in regions where
the equivalent Hermitian Hamiltonian $h$ is simply $p^2$. Otherwise\cite{HFJ-1df} the conservation of probability takes a non-local form involving an integral over $h(x,y)$.

\subsection{Effect of $K_\alpha K_\beta$}

 According to Eq.~(\ref{K}) we need to calculate $K_\alpha{Q_1}$.
 In general\cite{AM-eta}, the effect of $K_\alpha$ on $u(x,y)$ is
\bea\label{Ka}
(K_\alpha u)(x,y)=-\half\alpha\left[\theta(y)\int_{x-y}^{x+y} dr\, u(r,0)+\theta(x)\int_{y-x}^{y+x} ds\, u(0,s)\right]\ .
\eea
Recall, cf. Eq.~(\ref{eta}), that $Q_1(x,y)=\half i \ve(y-x)[\theta(x+y+2L)-\theta(x+y-2L)]$.
Thus
\bea\label{KaQ1}
(K_\alpha Q_1)(x,y)&=&\quarter i\alpha\theta(y)\int_{x-y}^{x+y} dr\,\ve(r)[\theta(r+2L)-\theta(r-2L)]\nonumber\\
&-& \quarter i\alpha\theta(x)\int_{y-x}^{y+x} ds\, \ve(s)[\theta(s+2L)-\theta(s-2L)].
\eea
In principle we need to calculate the effects of $K_\alpha^n$ arising from the expansion of
Eq.~(\ref{K}). However, a surprising and welcome result is that these vanish for $n>1$. Thus, in the
calculation of $K_\alpha^2 Q_1$ according to Eq.~(\ref{Ka}), we need
\bea
(K_\alpha Q_1)(r,0)&=&
- \quarter i\alpha\theta(r)\int_{-r}^{r} du\, \ve(u)[\theta(u+2L)-\theta(u-2L)]\nonumber\\
&=&0\;\; \mbox{by symmetry}.
\eea
Similarly $(K_\alpha\bar{Q}_1)(0,s)=0$.
So in fact all higher order terms in Eq.~(\ref{K})
are absent.

\subsubsection{Bound-state wave-function}
Here, since we are concerned with the limit as $L\to\infty$, we need to evaluate
Eq.~(\ref{KaQ1}) for  $0<x\ll L$. A similar analysis will apply to $x<0$, but this is easily
obtained by $PT$ symmetry. First it is easy to see that for a non-zero result $x$ and $y$
must have opposite signs. So in the present case $y<0$. There are then three possibilities depending
on the positions of the limits of integration $y+x$ and $y-x$.
The net result is
\bea\label{x>}
(K_\alpha{Q}_1)(x,y)&=&-\quarter i\alpha\left\{\begin{array}{lc}
\phantom{-}2y&\;\; 0>y>-x \\-2x & \;\; -x>y>-2L+x \\ -(2L+y+x) & \;\; -2L+x>y>-2L-x\end{array}\right.,
\eea
giving a correction to the bound-state wave function
\bea
\Delta\Psi&=&\frac{1}{8}i\alpha\beta\left\{2\int_{-x}^0 y\ -2x\int_{-2L+x}^{-x}
-\int_{-2L-x}^{-2L+x} (2L+y-x)\right\}e^{\kappa y}dy
\nonumber\\
&=&\half i \tilde{\alpha}\tilde{\beta}\left[e^{-2\kappa L}(e^{\kappa x}-e^{-\kappa x})-2(1-e^{-\kappa x})\right].
\eea
When added to $\Psi$ of Eq.~(\ref{Psibound}), and remembering that $\tilde{\alpha}=1+O(\beta^2)$,
we obtain the remarkably simple result
\bea\label{Psibcorr}
\Psi(x)=e^{-\kappa x} +O(\beta^2),
\eea
which to this order is equal to the original undisturbed wave function $\psi(x)$
of Eq.~(\ref{unp}) for $x>0$.

\subsubsection{Scattering wave-function}
For the corrections to the scattering wave-function we need to evaluate Eq.~(\ref{KaQ1}) for $|x|\gg L$.
Let us first consider $x\gg L$, in order to obtain the correction $\Delta\Psi_>$ to $\Psi_>$.
Again, for a non-zero result $y$ must be negative, so that
\bea
(K_\alpha Q_1)(x,y)=-\quarter i\alpha\int_{y-x}^{y+x} ds\, \ve(s)[\theta(s+2L)-\theta(s-2L)].
\eea
Here the lower limit, $y-x$, is less than $-2L$, and there are again two possibilities
depending on the position of the upper limit. The net result is
\bea
(K_\alpha\ Q_1)(x,y)=\quarter i\alpha\left\{
\begin{array}{lc} \phantom{-(}x+y-2L &\;\; 0<x+y<2L  \\ -(x+y+2L) &\;\; -2L<x+y<0 \\0 & \;\;\mbox{otherwise}\end{array}\right. ,
\eea
giving
\bea
\Delta\Psi_>&=&-\frac{1}{8}\alpha\beta\left[\int_{-2L-x}^{-x}dy(x+y-2L)\psi_<(y)\right.\nonumber\\
&&\left.\hspace{1cm}-\int_{-x}^{2L-x}dy(x+y+2L)\psi_<(y) \right]
\eea
where $\psi_<(y)=(1-i\hat{\alpha}) e^{iky}+i\hat{\alpha}e^{-iky}$.
The result of this calculation is
\bea
\Delta\Psi_>=2\hat{\alpha}\hat{\beta}\sin^2{kL}\left[(1-i\hat{\alpha})e^{-ikx} +i\hat{\alpha}e^{ikx}\right],
\eea
giving the corrected value of $\Psi_>$ as
\bea\label{Psi>corr}
\Psi_>(x)&=&(1-\hat{\alpha}\hat{\beta}\sin{2kL}+2\hat{\alpha}^2\hat{\beta}\sin^2{kL})e^{ikx}\nonumber\\
&&+i\hat{\beta}(1-i\hat{\alpha})(\sin{2kL}+2\hat{\alpha}\sin^2{kL})e^{-ikx}.
\eea
Now let us take $x\ll -L$ , so that $y$ must be positive. The expression for $K_\alpha Q_1$ turns
out to be the same as that given in Eq.~(\ref{x>}). Thus
\bea
\Delta\Psi_<&=&-\frac{1}{8}\alpha\beta\left[\int_{-2L-x}^{-x}dy(x+y-2L)\psi_>(y)\right.\nonumber\\
&&\left.\hspace{1cm}-\int_{-x}^{2L-x}dy(x+y+2L)\psi_>(y) \right]
\eea
where $\psi_>(y)=e^{iky}$. Hence
\bea
\Delta\Psi_<=2\hat{\alpha}\hat{\beta}\sin^2{kL}e^{-ikx},
\eea
giving the corrected value of $\Psi_<$ as
\bea\label{Psi<corr}
\Psi_<(x)=(1-i\hat{\alpha})e^{ikx}+i[\hat{\alpha}+\hat{\beta}\sin{2kL}-2\hat{\alpha}\hat{\beta}\sin^2{kL})e^{-ikx}.
\eea
From Eqs.~(\ref{Psi>corr}) and (\ref{Psi<corr}) we obtain the fluxes
\bea
\Phi_>=\Phi_<=1-2\hat{\alpha}\hat{\beta}(\sin{2kL}-2\hat{\alpha}\sin^2{kL})+O(\beta^2).
\eea
Compared with Eq.~(\ref{firstfluxes}), we now have the explicit $O(\alpha^2\beta)$ terms, and the
result is correct to the order shown.
\section{Discussion}
\setcounter{footnote}{1}
\renewcommand{\thefootnote}{{\fnsymbol{footnote}}}

Let us now consider the conceptual issues raised by these calculations. First it should be emphasized that we are concerned
 here with quasi-Hermitian Hamiltonians, that is, Hamiltonians that can be related by a similarity transformation to a Hermitian
 Hamiltonian. It is only for such Hamiltonians that we can attempt to construct the metric $\eta$ and to look at the situation in
 the quasi-Hermitian framework. For generic non-Hermitian Hamiltonians no such framework is available, and one is bound to treat
 them as effective Hamiltonians within the standard framework of quantum mechanics. In that case one would simply perform an
 analysis similar to that of Section 2 and accept that unitarity is not conserved, essentially because we are dealing with a
 subsystem of a larger system whose physics has not been taken fully into account.

For the potential we have chosen we are able to consider both bound and scattering states. The bound state is the simpler to consider.
The standard quantum mechanical analysis shows that the introduction of the perturbing $PT$-symmetric delta-function potentials does
not significantly modify either the bound-state energy or the wave-function if these potentials are sufficiently distant. The analysis of
Section 3, in particular the final result of Eq.~(41), shows that this remains the case within the quasi-Hermitian framework, with
the new bound-state wave-function $\Psi(x)$ being identical to the original wave-function $\psi(x)$ for large $L$.
This is a new and reassuring result, which we were unable to address in Ref.~\cite{HFJ-1df}, where the potential did not support
a bound state. The general message we would like to draw from this is that a localized physical Hermitian
system is not significantly affected by the introduction of distant non-Hermitian potentials, and may be treated
in the framework of standard quantum mechanics without the necessity of introducing a new metric.

The real conceptual problems arise for the scattering states. The standard quantum mechanical analysis shows that unitarity,
as conventionally defined, is not conserved. In the quasi-Hermitian framework, however, one can hope that a modified form of
unitarity is in fact conserved, as is indeed borne out by the calculations of Section 3. However, this involves a fairly drastic
redefinition of asymptotic states, a non-local effect, given the finite support of the perturbing $PT$-symmetric potentials.
The crux is the difference between $x$, the coordinate parameter in terms of which the Hamiltonian is originally
defined, and $X$, its quasi-Hermitian counterpart, defined by the non-local relation $X\equiv \rho x \rho^{-1}$. The former,
$x$, is Hermitian and therefore an observable, in the standard framework of quantum mechanics, while $X$ is not. Conversely,
in the quasi-Hermitian picture, $X$ is quasi-Hermitian and an observable, while $x$ is not. The argument of the first wave-function
$\psi(x)$ is the eigenvalue of $x$, while that of $\Psi(x)$ is the eigenvalue of $X$.

What we have done in this paper is transform an initial scattering set-up defined in terms of $x$, and then consider the
corresponding picture in the quasi-Hermitian framework in terms of $X$. The initial scattering set-up had plane waves entering
from the left and then being either reflected or transmitted, with probability not being conserved. As we have seen, the
corresponding quasi-Hermitian picture is that the newly-defined probability is indeed conserved, but that
waves now enter from both left and right. An alternative mathematical possibility would be to set up a scattering situation
in which $\Psi_>$ has only outgoing waves and then work backwards to construct $\psi$, which would undoubtedly have waves
entering from both left and right. In either case, the physical picture changes drastically when going from one picture
to the other\footnote{In a recent paper, ref.~{\cite{MZ}}, Znojil has constructed certain discrete matrix models of scattering in which
in and out states are not mixed in the quasi-Hermitian picture, but instead the definition of the flux differs on the left and
right of the scattering centre. This is a somewhat less drastic change, but still involves a departure from standard quantum
mechanics at large distances.}.

In the author's opinion, the only satisfactory resolution of this dilemma is to treat the non-Hermitian scattering potential as an
effective one, and work in the standard framework of quantum mechanics, accepting that this effective potential may well involve
the loss of unitarity when attention is restricted to the quantum mechanical system itself and not its environment. This is indeed
the attitude taken in a recent paper by Berry \cite{MBerry}, where various intensity sum rules are derived for diffraction off
$PT$-symmetric optical lattices. There it is taken for granted that the intensities are given by $|\psi|^2$. It is true that
these are classical calculations, but because of the correspondence principle the same thing would apply in quantum mechanics.

\end{document}